\documentclass[doublecol,linenumbers,figures]{epl2}
\usepackage{amsfonts,amsmath,amssymb}
\usepackage{color,soul}
\institute{
  \inst{1}
 Department of Applied Mathematics, School of Mathematics, University of Leeds, Leeds LS2 9JT, UK\\
  \inst{2} 
Department of Physics \& Astronomy, Iowa State University, Ames, IA 50011, USA\\
  \inst{3} 
Department of Physics, Virginia Polytechnic Institute \& State University, Blacksburg, VA 24061, USA
  }
\pacs{89.75.-k}{Complex systems}
\pacs{02.50.-r}{Probability theory, stochastic processes, and statistics}
\pacs{05.40.-a}{Fluctuation phenomena, random processes, noise, and Brownian motion}

\abstract{We introduce an heterogeneous nonlinear $q$-voter model with zealots and two types of susceptible voters, and study its {\it non-equilibrium} properties when the population is finite  and well mixed.
In this two-opinion model, each individual supports one of two parties and is either a zealot or a susceptible voter of type $q_1$ or $q_2$. While here zealots never change their opinion, a $q_i$-susceptible voter ($i=1,2$) consults a group of $q_i$ neighbors at each time step, and adopts their opinion if all group members agree.
We show that this model violates the detailed balance whenever $q_1 \neq q_2$ and has surprisingly rich properties. 
Here, we focus on the characterization of the model's non-equilibrium stationary state (NESS) in terms of its 
probability distribution and currents in the distinct regimes of low and high density of zealotry. We unveil the
NESS properties in each of these phases by computing the opinion distribution and the circulation of probability currents, as well as the  two-point correlation functions at unequal times (formally related to a ``probability angular momentum''). Our analytical calculations obtained in the realm of a linear Gaussian approximation are compared with numerical results.}
\input{tcilatex}

\begin{document}

\title{Characterization of the Nonequilibrium Steady State of a \\
Heterogeneous Nonlinear $q$-Voter Model with Zealotry}
\author{Andrew Mellor \inst{1} \and Mauro Mobilia \inst{1} \and R.K.P. Zia %
\inst{2,3}}

\shorttitle{Nonlinear $q$-voter model: heterogeneity \& zealotry}
\shortauthor{A. Mellor \etal}

\maketitle

\section{Introduction}

Since Schelling's pioneering work~\cite{Schelling} there has been increasing
interest in using simple theoretical models to describe social phenomena
such as the dynamics of opinions~\cite{Opinions}. In this context,
individual-based models commonly used in statistical physics are
particularly insightful, as they reveal the micro-macro connection in social
dynamics~\cite{Schelling,Opinions}. The voter model (VM)~\cite{Liggett}
serves as a reference to describe the evolution of opinions in socially
interacting populations. (See \textit{e.g.}~\cite{Opinions,Sociophysics}
and
references therein.) In spite of its paradigmatic role, the VM relies on a
number of unrealistic assumptions, such as the total lack of self-confidence
of all voters and their perfect conformity, which invariably leads to a
consensus. In fact, it has been shown that members of a society respond
differently to stimuli and this greatly influences the underlying social
dynamics~\cite{Granovetter,ConfIndep,GroupSize}.
An approach to model a
population with different levels of confidence is to assume that some agents
are \textquotedblleft zealots\textquotedblright\ who favor one opinion~\cite%
{MM1} 
or maintain a fixed opinion~\cite{zealot07}. Since the introduction of
these simple types of behavior in the VM, a large variety of zealot models
have been studied, see, {\it e.g.},~Refs.~\cite{otherZealots}.

In this Letter, we focus on a variant of the VM known as the two-state
nonlinear $q$-voter model ($q$VM)~\cite{qVM} which has attracted much
interest~\cite{exitprobq}.
In the $q$VM, a voter can be influenced by a
group of $q$ neighbors. The version with $q=2$ is closely related to the
well-known models of Refs.~\cite{Sznajd}.
Motivated by important psychology
and sociology tenets~\cite{ConfIndep,Granovetter}, the basic ideas
underlying the $q$VM and zealotry have been combined into the $q$-voter
model with \textit{inflexible} zealots ($q$VMZ)~\cite{MM15}. Indeed, social
scientists have established that conformity by imitation, an important
mechanism for collective actions, is observed only when the group-size is
large and can be altered by individuals who are able to resist the group
pressure~\cite{GroupSize,ConfIndep}. It is understandable that social
conformity is unlikely for small groups and can be significantly suppressed
by the presence of zealots. In the $q$VMZ dynamics, both group-size limited
conformity and zealotry are accounted for. Furthermore, in a well-mixed
setting, this dynamics obeys detailed balance, so that the exact stationary
distribution is easily found~\cite{MM15}. The system resembles one in
thermal equilibrium, characterized by two phases: As zealotry is lowered
through a critical level, the opinion distribution transitions from being
single-peaked to being bimodal, with non-trivial switching dynamics (in
finite populations).

Unlike in the models described above, the populations in our society are
highly heterogeneous. In principle, we may describe the relevant situation
of different responses to social stimuli by considering a distribution of $q$%
's~\cite{Granovetter}. Does this generalization modify the conclusions of
the $q$VMZ? If so, how? In this Letter, we explore the simplest way to this
broader view, namely, a population with just two subgroups with different $q$%
's (denoted by $2q$VZ): $q_{1}<q_{2}$. Similar to the $q$VMZ, phase
transitions exist in the $2q$VZ. Unlike in $q$VMZ, the dynamics of this
model does not obey detailed balance, so the system relaxes into a
non-equilibrium steady state (NESS). Thus, from the standpoint of
statistical physics, $2q$VZ is a highly non-trivial extension. In general,
there is no simple way to compute a NESS distribution~\cite{Hill66}, while
persistent probability current prevails~\cite{ZS07}. As a consequence, there
are observable quantities which are trivially zero in the $q$VMZ that do 
\textit{not} vanish in the $2q$VZ. For example, opinions among those with
smaller $q$ change more readily, and \textquotedblleft
drive\textquotedblright\ those in the other subgroup. Though subtle,
directed \textquotedblleft oscillations\textquotedblright\ associated with
fluctuating quantities can be measured. Here, we report results of a
baseline study with the simplest case $q_{1,2}=1,2$. Our methods include
stochastic simulations and numerical solutions of the master equation for
systems finite size $N$, as well as a continuum version based on the
Fokker-Planck equation and its analysis through a linear Gaussian
approximation (LGA)~\cite{Lax66,JBW03,JBW07,ZS07}.

\section{Model specification and NESS \label{sec:model}}

Our model, the $2q$VZ, consists of a population of $N$ voters who support
one of two parties, the opinion of each denoted by $\pm 1$. Some voters are
inflexible zealots, never changing their opinions. Their numbers are denoted
by $Z_{\pm }$, the subscript showing the associated opinion. The rest are
swing voters of two types, denoted by $q_{1}$ and $q_{2}$. Known as $q_{i}$%
-susceptibles, their numbers are $S_{i}$, with $i=1,2$. During the
evolution, each agent maintains its behavior, so that $Z_{\pm }$ and $S_{i}$
are all conserved (with $S_{1}+S_{2}+Z_{+}+Z_{-}=N$). At each time step, a
voter is chosen at random and if it is a zealot then no action is taken.
However, if a $q_{i}$-susceptible is chosen, then it collects the opinions
from a random group of $q_{i}$ neighbors and adopts the opinion of the group
only if their opinion is \textit{unanimous}~\footnote{
As in Refs.~\cite{qVM,exitprobq,MM15}, we allow repetition. Note that in
Ref.~\cite{qVM} a voter can change its opinion with a flip rate $\epsilon $
even in the absence of consensus among its $q$ neighbors. Here, as in most
of Refs.~\cite{exitprobq,MM15}, we set $\epsilon =0$.} (see supplementary material (SM) \cite{AMZ_SM}). For simplicity, we
investigate this model on a complete graph (well-mixed population). Since
there is no spatial structure, our system is completely specified by the
number $n_{i}$ of $q_{i}$-susceptible voters holding opinion $+1$
(also denoted by $\vec{n}=\left(n_{1},n_{2}\right) $).

Since configuration space is a discrete set of $S_{1}\times S_{2}$ points
and updates involve a single step to a nearest neighbour on a square
lattice, our system behaves exactly as a two-dimensional random walker, with
inhomogeneous and biased rates. Thus, our simulation runs consist of
recording the trajectories of such a random walker. Meanwhile, the full
stochastic process is defined by a master equation (ME) for the evolution of 
$P\left( \vec{n};T\right) $~\cite{noise},
 the probability to find our system
in state $\vec{n}$, $T$ time steps (attempts) after some initial
configuration $\vec{n}_{0}$. Since our main interest is the stationary
distribution, $P^{\ast }\left( \vec{n}\right) $, we suppress references to $%
\vec{n}_{0}$. As $T$ increases by unity, a walker at $\vec{n}$ steps to $%
\vec{n}^{\prime }$ with probability $W\left( \vec{n}\rightarrow \vec{n}%
^{\prime }\right) $, a process represented by the ME $P\left( \vec{n}%
;T+1\right) =\sum_{\vec{n}^{\prime }}\mathcal{G}\left( \vec{n},\vec{n}%
^{\prime }\right) P\left( \vec{n}^{\prime };T\right) $. Here, $\vec{n}%
^{\prime }\in \left\{ (n_{1}\pm 1,n_{2}),\left( n_{1},n_{2}\pm 1\right)
\right\} $ is one of the four nearest neighbors of $\vec{n}$, from which the
transitions occur with respective stepping probabilities $W_{1}^{\pm }(\vec{n%
})$ and $W_{2}^{\pm }(\vec{n})$: 
\begin{eqnarray}
W_{i}^{+}(\vec{n}) &=&\frac{S_{i}-n_{i}}{N}\left( \frac{Z_{+}+n_{1}+n_{2}}{%
N-1}\right) ^{q_{i}}  \label{w+} \\
W_{i}^{-}(\vec{n}) &=&\frac{n_{i}}{N}\left( \frac{%
Z_{+}+S_{1}+S_{2}-n_{1}-n_{2}}{N-1}\right) ^{q_{i}}.  \label{w-}
\end{eqnarray}%
From here, it is straightforward to write an explicit form for $\mathcal{G}$%
, as well as joint probabilities $\mathcal{P}\left( \vec{n},T;\vec{n}%
^{\prime },T^{\prime }\right) $ at two different times\footnote{%
Assuming $T>T^{\prime }$, $\mathcal{P}\left( \vec{n},T;\vec{n}^{\prime
},T^{\prime }\right) =\mathcal{G}^{T-T^{\prime }}\left( \vec{n},\vec{n}%
^{\prime }\right) P\left( \vec{n}^{\prime },T^{\prime }\right) $. Details
will be provided in \cite{AMZ2}.}. Much of our attention here will be
devoted to the change, $P\left( \vec{n};T+1\right) -P\left( \vec{n};T\right) 
$, given by a sum of \textit{probability currents} which account for
transitions into, or out-of, the configuration $\vec{n}$. Specifically, the 
\textit{net} probability current from $\vec{n}$ to $\vec{n}^{\prime }\equiv
(n_{1}+1,n_{2})$ is $K_{1}(\vec{n};T)=W_{1}^{+}(\vec{n})P\left( \vec{n}%
;T\right) -W_{1}^{-}(\vec{n}^{\prime })P\left( \vec{n}^{\prime };T\right) $,
and a similar expression for $K_{2}(\vec{n};T)$ for $\vec{n}$ to $%
(n_{1},n_{2}+1)$. Thus, $\mathcal{G}P$ is intimately related to the current $%
\vec{K}=(K_{1},K_{2})$.

To verify that this dynamics violates detailed balance (and time reversal),
we may apply the Kolmogorov criterion~\cite{Kol36} on any closed loop, the
simplest being four $\vec{n}$'s around a plaquette~\cite{AMZ2}. As a
consequence, our system settles into a NESS, with non-trivial $P^{\ast }$
and stationary current $\vec{K}^{\ast }$. Though the main behavior of our
model is qualitatively the same as in the $q$VMZ, the presence of $\vec{K}%
^{\ast }$ leads to important, distinguishing features, displayed through
physical observables, such as means, $\left\langle n_{i}\right\rangle \equiv
\sum_{\vec{n}}n_{i}P^{\ast }\left( \vec{n}\right) $, and correlations, $%
\left\langle n_{i}n_{j}\right\rangle _{T}\equiv \sum_{\vec{n},n^{\prime
}}n_{i}^{\prime }n_{j}\mathcal{P}^{\ast }\left( \vec{n},T;\vec{n}^{\prime
},0\right) $. Note that the order of indices in the latter is crucial: $i$ ($%
j$) is associated with the earlier (later) variable when $T>0$. One key
observable is the \textit{antisymmetric} part of $\left\langle
n_{i}n_{j}\right\rangle _{T\neq 0}$: Being odd under time reversal, it
highlights the underlying NESS characteristic, and so, vanishes in the $q$%
VMZ. When the time difference is infinitesimal ($T=1$, large $N$), it can be
identified as the total \textquotedblleft probability angular
momentum\textquotedblright $~$\cite{SZ2014}, in analogy with the classical
angular momentum associated with current loops in fluids. To illustrate
these novel features, we examine in detail a specific model -- $Z_{+}=Z_{-}$%
, $S_{1}=S_{2}$, $q_{1,2}=1,2$ -- and provide several explicit results below.

\section{Analytic results \& simulation studies \label{results}}

\hspace{0.1pt}\newline
\textit{Mean-field Approximation (MFA):} This approach offers the most
intuitive picture, valid when $N\rightarrow \infty $ with fixed densities: $%
\left( z_{\pm },s_{i},x_{i}\right) =\left( Z_{\pm },S_{i},n_{i}\right) /N$
(continuous variables subjected to $z_{+}+z_{-}+s_{1}+s_{2}=1$ and $x_{i}\in %
\left[ 0,s_{i}\right] $). In this limit, the rates $W_{i}^{\pm }$ become $%
w_{i}^{+}(\vec{x})=(s_{i}-x_{i})\mu ^{q_{i}}$ and $w_{i}^{-}(\vec{x}%
)=x_{i}(1-\mu )^{q_{i}}$, where $\mu \equiv z_{+}+x_{1}+x_{2}$ clearly
represents the fraction holding opinion $+1$. In the MFA, averages of
products are replaced by products of averages, and from the ME, we find the
rate equations (REs) 
\begin{equation}
\dot{x_{i}}\equiv \partial _{t}x_{i}=w_{i}^{+}-w_{i}^{-}=(s_{i}-x_{i})\mu
^{q_{i}}-x_{i}(1-\mu )^{q_{i}}.  \label{RE}
\end{equation}%
As in the $q$VMZ\cite{MM15}, the REs admit one or three fixed points,
depending on $z_{\pm }$. At the fixed point(s), $x_{i}^{\ast }$, the ratio $%
\rho \equiv \mu ^{\ast }/\left( 1-\mu ^{\ast }\right) $ satisfies $%
z_{+}+\sum_{i=1,2}s_{i}/(1+\rho ^{q_{i}})=1/(1+\rho )$, while $x_{i}^{\ast
}=s_{i}/(1+\rho ^{q_{i}})$. For models with $z_{+}=z_{-}$, $\rho =1$ is
always a solution, associated with the symmetric fixed point $x_{i}^{\ast
}=s_{i}/2$ (denoted by $\vec{x}^{(0)}$ below). Above a critical density of
zealots, $z_{c}$, this is the only fixed point and is stable. For $z<z_{c}$, 
$\vec{x}^{(0)}$ turns unstable, while two others (denoted by $\vec{x}^{(\pm
)}$) emerge and both are stable.

For our specific model ($z_{\pm }=z$, $s_{i}=s=1/2-z$, $q_{i}=i$), the MFA
for $z_{c}$ is $1/6$ (i.e., $s_{c}=1/3$). When $z<z_{c}$, $\vec{x}^{(\pm )}$
are given by the two solutions to $\rho +(1/\rho )+1=1/(2z)$, associated
with the spontaneously breaking of the Ising-like symmetry ($\rho
\Leftrightarrow 1/\rho $). Thus, this MFA predicts the same phase transition
as in the $q$VMZ (pitchfork bifurcation~\cite{MM15}). Of course, being
deterministic, it cannot account for fluctuations. \vspace{0.1cm} \newline
\textit{Fokker-Planck equation (FPE):} In finite populations, demographic
fluctuations are important, as they drive interesting time-dependent
phenomena in the stationary state. For large but finite $N$, these
fluctuations are embodied in the \textit{probability density} $P\left( \vec{x%
};t\right) $, the continuum version of $P\left( \vec{n};T\right) $. Here, $%
t\equiv T/N$ also becomes continuous, as $P\left( \vec{n};T+1\right)
-P\left( \vec{n};T\right) \rightarrow N^{-1}\partial _{t}P\left( \vec{x}%
;t\right) $. To the leading, non-trivial order in $1/N$, the evolution is
adequately captured by the FPE: $\partial _{t}P\left( \vec{x};t\right)
=\sum_{i=1,2}\frac{\partial }{\partial x_{i}}\left[ \frac{\partial }{%
\partial x_{i}}u_{i}(\vec{x})P+v_{i}(\vec{x})P\right] $~\cite{noise}, where $%
u_{i}\equiv \left( w_{i}^{+}+w_{i}^{-}\right) /2N$ and $v_{i}\equiv
w_{i}^{-}-w_{i}^{+}$. Clearly, the right-hand-side can be identified as the
divergence of the \textit{probability current density} 
\begin{equation*}
K_{i}(\vec{x};t)=-\partial \left[ u_{i}P\right] /\partial x_{i}-v_{i}P
\end{equation*}%
(continuum version of $K_{i}(\vec{n};T)$). The stationary probability
density $P^{\ast }\left( \vec{x}\right) $ is given by $0=\vec{\nabla}\cdot 
\vec{K}^{\ast }$. For a dynamics which satisfies detailed balance, $\vec{K}%
^{\ast }$ necessarily vanishes, leading to equations for $P^{\ast }$ that
can be easily solved~\cite{noise}. 
In a NESS, $\vec{K}^{\ast }\neq 0$, i.e., non-trivial currents persist.
Clearly, the curl of $\vec{K}^{\ast }$ does not vanish and, known in fluid
dynamics as the vorticity, $\vec{\nabla}\times \vec{K}^{\ast }$ is a
one-component field in two dimensions. Of course, $\vec{K}^{\ast }$ can also
be expressed as the curl of another field, the stream function. (See~\cite{AMZ_SM} for some details). In other words, 
the currents form closed loops, which lead us to identify 
\begin{equation}
L_{ij}=\int_{\vec{x}}\left[ x_{i}K_{j}^{\ast }(\vec{x})-x_{j}K_{i}^{\ast }(%
\vec{x})\right] ~d{\vec{x}}  \label{Lij}
\end{equation}%
as the total `\textit{probability angular momentum}'~\cite{SZ2014}, by
formal analogy with the total mass angular momentum ($\int_{\vec{x}}\vec{x}%
\times \vec{J}~d{\vec{x}}$) in fluids with current density $\vec{J}$. As a
pseudotensor in arbitrary dimensions, $L_{ij}$ has just a single independent
component, (say) $L_{12}=\mathcal{L}$. Since $\vec{K}^{\ast }$ is linear in $
P^{\ast }$, we identify $\mathcal{L}$ as the steady state average of a
function of $\vec{x}$. Below we show that the simple approximation $\vec{K}
^{\ast }\propto \vec{x}P^{\ast }$ provides many analytic results, {\it e.g.}, an
 expression for the two-point correlation at \textit{unequal} times \footnote{
Also known as the lagged correlation or lagged covariance.} 
\begin{eqnarray}
C_{ij}\left( \tau \right) \equiv \left\langle x_{i}x_{j}\right\rangle _{\tau
} \quad (\text{for} \quad \tau\neq 0).  \label{Cij}
\end{eqnarray}
Decomposing $C_{ij}$ into the symmetric and antisymmetric parts, we see that 
$\tilde{C}_{ij}\equiv C_{ij}-C_{ji}$ is \textit{odd} in $\tau $,
highlighting time reversal violation and serving as a principal
characteristic of a NESS. Indeed, to lowest order (in $\vec{x}$), $L_{ij}$
is given by $\left. \partial _{\tau }\tilde{C}_{ij}\right\vert _{0}$. 
\vspace{0.1cm} \newline
\textit{Linear Gaussian Approximation (LGA):} In this scheme, we consider 
\textit{deviations} from a fixed point, $\vec{\xi}\equiv \vec{x}-\vec{x}%
^{\ast }$, and, keeping the lowest non-vanishing order, we find the
linearized version of the FPE~\cite{noise} 
\begin{equation}
\partial _{t}P(\vec{\xi},t)=\sum_{i,j}\partial ^{i}\left\{ D_{ij}\partial
^{j}+F_{i}^{j}~\xi _{j}\right\} P(\vec{\xi},t),  \label{LGM}
\end{equation}%
where $\partial ^{i}\equiv \partial /\partial \xi _{i}$, $D_{ij}=\delta
_{ij}w_{i}^{\ast }/N$ with $w_{i}^{\ast }=w_{i}^{+}\left( \vec{x}^{\ast
}\right) =w_{i}^{-}\left( \vec{x}^{\ast }\right) $, and $F_{i}^{j}\equiv
-(\partial \dot{x_{i}}/\partial x_{j})|_{\vec{x}=\vec{x}^{\ast }}$ is the
linear stability matrix at $\vec{x}^{\ast }$. Thus, the LGA is defined by
two matrices: $D_{ij}$ and $F_{i}^{j}$, or $\mathbb{D}$ and $\mathbb{F}$ for
short. This linearized FPE can be translated into Langevin equations with
linear drift $-\mathbb{F}\vec{\xi}$ plus Gaussian white noise controlled by $%
\mathbb{D}$~\cite{noise}. Below we show that the LGA provides much insight
into the non-equilibrium character of our model by allowing us to find
analytic expressions for various quantities. From the explicit expressions~%
\cite{AMZ_SM}, we find that $\det \mathbb{F}\propto 1-6z$ (in all cases) and
its eigenvalues, denoted by $\lambda _{\pm }$, are positive in the regions
of interest. Also expected, $\mathbb{D}$ is $O\left( 1/N\right) $, so that
fluctuations of $\vec{\xi}$ are $O\left( 1/\sqrt{N}\right) $, and we note
that $D_{11}>D_{22}$ which confirms that the $q_{1}$-susceptibles are more
likely to change their opinions.

If $\mathbb{D}^{-1}\mathbb{F}$ is symmetric, then detailed balance is
satisfied and $P^{\ast }\propto \exp \left\{ -\vec{\xi}\mathbb{D}^{-1}%
\mathbb{F}\vec{\xi}/2\right\} $ is a Gaussian distribution for which $%
\mathbb{D}^{-1}\mathbb{F}/2$ is the \textquotedblleft
potential\textquotedblright . Here, we find that $\mathbb{D}^{-1}\mathbb{F}$
is \textit{not} symmetric. Nevertheless, the solution of (\ref{LGM}) is
still a Gaussian~\cite{Lax66,JBW03,JBW07,ZS07}: $P^{\ast }( \vec{\xi}%
) \propto \exp \left( -\vec{\xi}\mathbb{C}^{-1}\vec{\xi}/2\right) $,
where the elements of $\mathbb{C}$\ are $\left\langle \xi _{i}\xi
_{j}\right\rangle _{0}$, the truncated $C_{ij}\left( 0\right) $\footnote{$%
\left\langle \xi _{i}\xi _{j}\right\rangle _{0}\equiv \langle
x_{i}x_{j}\rangle _{0}-\langle x_{i}\rangle \langle x_{j}\rangle
=C_{ij}\left( 0\right) -\langle x_{i}\rangle \langle x_{j}\rangle $.}. It
can be expressed in terms of the eigenvectors and eigenvalues of $\mathbb{F}~
$\cite{Lax66,ZS07}, or by solving $\mathcal{S}\left[ \mathbb{FC}\right] =%
\mathbb{D~}$\cite{JBW03,JBW07}, where $\mathcal{S}\left[ \mathbb{FC}\right] $
denotes the \textit{symmetric part} of $\mathbb{FC}$. Deferring the explicit
forms of $\mathbb{C}$ to elsewhere$~$\cite{AMZ2}, we present here the
implications for $\vec{K}^{\ast }$ and the physical observables.

Since $\vec{\xi}P^{\ast }=-\mathbb{C}\vec{\partial}P^{\ast }$, the steady
currents in Eq.~(\ref{LGM}) can be written as $\vec{K}^{\ast }=\left[ 
\mathbb{FC-D}\right] \vec{\nabla}P^{\ast }$~\cite{ZS07}. We emphasize that,
given $\mathcal{S}\left[ \mathbb{FC}\right] =\mathbb{D}$, the matrix $%
\mathbb{FC-D}$ is precisely the \textit{antisymmetric part} of $\mathbb{FC}$
, here denoted by $\mathcal{A}\left[ \mathbb{FC}\right] $. Thus, $\vec{K}
^{\ast }$ is manifestly divergence free.

Proceeding to observables, we consider $L_{ij}$ or simply, $\mathbb{L}$.
Since $\int_{\vec{x}}\vec{K}^{\ast }d\vec{x}=0$, the $\vec{x}$ in Eq.~(\ref%
{Lij}) can be replaced by $\vec{\xi}$. In the framework of the LGA, we
readily find the remarkably simple expression $\mathbb{L}=2\mathcal{A}\left[ 
\mathbb{FC}\right] $. In this setting, we see that $\mathbb{FC}=\mathbb{D}+%
\mathbb{L}/2$, placing this angular momentum on an equal footing with
diffusion\footnote{%
Note our distribution has unit \textquotedblleft mass\textquotedblright\ ($%
\int P^{\ast }=1$) so that the units of $\mathbb{L}$ are $%
x^{2}/t$, precisely those of diffusion.}. In our specific $2q$VZ, the only
independent component is $\mathcal{L}$, given by $s^{2}/\left[ \left(
6-8s\right) N\right] $ and \ $4z^{2}/\left[ \left( 2z+3\right) N\right] $
for $\vec{x}^{(0)}$ and $\vec{x}^{(\pm )}$, respectively. Further, we recall
that $\mathbb{L}$ is intimately related to the two-point correlation $%
\mathbb{C}\left( \tau \right) \equiv \left\langle \xi _{i}\xi
_{j}\right\rangle _{\tau }\footnote{%
Specifically, $\mathbb{L}=2\mathcal{A}\left[ \left. \partial _{\tau }\mathbb{%
\ C}\left( \tau \right) \right\vert _{0}\right] $.}$. In the LGA, $\mathbb{C}%
\left( \tau \right) $ is explicitly $\mathbb{C}\exp \left( -\mathbb{F}%
^{T}\tau \right) $ ~\cite{JBW07}. Here, its antisymmetric part is just one
independent quantity and so, we focus on (say) $\tilde{C}_{12}\left( \tau
\right) $. Dropping the subscript, we find the explicit expression%
\begin{equation}
\tilde{C}\left( \tau \right) =\mathcal{L}\left( \frac{e^{-\lambda _{-}\tau
}-e^{-\lambda _{+}\tau }}{\lambda _{+}-\lambda _{-}}\right)   \label{Ltau}
\end{equation}%
which exposes some noteworthy features: Clearly, $\left. \partial _{\tau }%
\tilde{C}\left( \tau \right) \right\vert _{0}=\mathcal{L}$. Moreover, its
long time behavior is governed by $\lambda _{-}$, the smallest eigenvalue of 
$\mathbb{F}$. Unlike typical correlations at unequal times, $\tilde{C}\left(
\tau \right) $ is non-monotonic, with a peak at $\hat{\tau}=\left( \lambda
_{+}-\lambda _{-}\right) ^{-1}\ln \left( \lambda _{+}/\lambda _{-}\right) $.

\begin{figure}[tbp]
\begin{center}
\includegraphics[width=3in, height=3in,clip=]{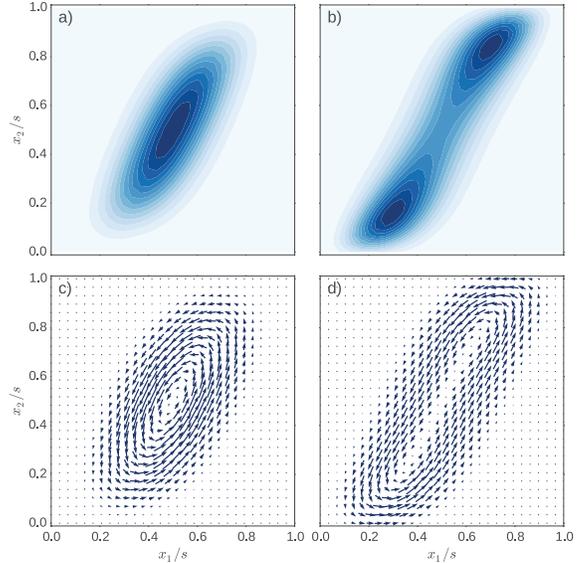}
\end{center}
\par
\vspace{-3mm}
\caption{\textit{(Color online)}. Stationary distribution and probability
current obtained from the exact numerical solution of the ME in the high/low
zealotry phase (left/right panels) with $S=30$. Top: $P^*$ as function of $%
\vec{x}=\vec{n}/N$; dark/light blue (grey) encodes a high/low probability.
Bottom: The wind-field vectors represent $\vec{K}^*$ at each point $\vec{x}$%
. Parameters are $Z=20$ and $N=100$ in (a, c), and $Z=13$ and $N=86$ in (b,
d). }
\label{Fig1}
\end{figure}

Below we show that all these predictions by the LGA are borne out in
simulations and exact numerical results (for  appropriate regions). Since
the LGA is formulated around a single fixed point, it clearly cannot
describe double-peaked distributions or escape times. By contrast, near each
peak, the reliability of the LGA improves in the limit of $N\rightarrow
\infty $ with fixed $z\neq 0,z_{c}$.\newline
\textit{Exact numerical solution:} For systems with small $S$, numerical
methods can be used to obtain $P^{\ast }$, by exploiting the relation $\left[
\mathcal{G}\left( \vec{n},\vec{n}^{\prime }\right) \right] ^{\infty
}=P^{\ast }\left( \vec{n}\right) $ (independent of $\vec{n}^{\prime }$!).
For example, for $S=30$, by iterating $\mathcal{G}^{2\tau }=\mathcal{G}%
^{\tau }\mathcal{G}^{\tau }$ just 64 times, we find changes at $\lesssim
10^{-20}$. Illustrated in Fig.~\ref{Fig1}(a,b) are heat maps (contour plots)
on $30\times 30$ grids for two cases: $Z=20$ and $13$. Associated with above
and below $z_{c}$, they clearly show the expected single \textit{vs.} double
peaked distributions. In the SM~\cite{AMZ_SM}, we show this transition in $%
S=50$ systems with many such plots compiled into a movie.

In Fig.~\ref{Fig1}a, we see that the contours resemble ellipses typical of
Gaussian distributions. Two other prominent features are: (i) The width in $%
n_{1}$ is much smaller than that in $n_{2}$, indicative of the relative ease
with which the $q_{1}$-susceptibles change their opinion, so that they stay
closer to the \textquotedblleft center\textquotedblright\ typically. Since
these widths scale with $\sqrt{N}$ ($\approx 10$ here), we should expect
only a qualitative fit from Gaussians. (ii) The alignment of the ellipses
implies a strong correlation between the two variables, an expected result
of both groups of susceptibles following the whims of the other.

In the $z<z_{c}$ case, we see that the $q_{2}$-susceptibles display a larger
spontaneous symmetry breaking than $q_{1}$-voters, reflecting the same
behavior as in the $q$VMZ. Meanwhile, if left alone, the $q_{1}$%
-susceptibles would reach a coexistence state~\cite{zealot07}. Thus, the
broken symmetry in the opinion of the former acts as an external imbalance
on the latter, dragging them to lean toward one pole or the other. Now, due
to the strong finite size effects, the region around each peak appears quite
asymmetric and seriously non-Gaussian when $N$ is small. We also note that
the peaks are linked via a \textquotedblleft ridge\textquotedblright
\thinspace\ , the lowest point along which is often referred to as a
\textquotedblleft gap\textquotedblright\ by mountaineers. If we consider $%
-\ln P^{\ast }$, then this gap represents the lowest barrier between two
\textquotedblleft wells.\textquotedblright\ Also known as the saddle point,
its height is expected to control the escape times faced by the random
walker trapped in one or the other well. As we expect the height difference
between the saddle and the well bottom to scale with $N$, we anticipate
escape times to scale with $e^{N}$, as found in the $q$VMZ~\cite{MM15}.
Finally, note that the analysis of the critical region $z\cong z_{c}$
necessitates a detailed finite size scaling study, which is beyond the scope
of this Letter.

From $P^{\ast }$, we have computed numerically  other exact quantities of
interest. The wind-field like plots of $\vec{K}^{\ast }$ shown in Fig.~\ref%
{Fig1}(c,d) provide good impressions of the general counterclockwise swirl.
Other characteristics (vorticity and stream function) are displayed in the
SM~\cite{AMZ_SM}. More quantitatively, the physical observables are readily
obtained and can be compared with the predictions of the LGA. Here, due to
symmetry, the exact $\left\langle \vec{x}\right\rangle $ is $\vec{s}/2$,
regardless of the location of the peak(s). The simplest non-trivial
quantities are two point correlations ($\mathbb{C},\mathcal{L}$) and we find
that the predictions of the LGA are in qualitative agreement with exact
calculations~\cite{AMZ_SM} in the high zealotry phase $z>z_{c}$. In
principle, we access the exact decay constants ($\lambda _{\pm }$) by
analyzing $\mathcal{G}^{\tau }$ numerically, a task deferred to a future
study~\cite{AMZ2}. In the low zealotry phase, $z<z_{c}$, we emphasize that $%
P^{\ast }$ includes \textit{all} trajectories, with visits near both $%
x^{\left( \pm \right) }$. Yet, the LGA can be expected to be good only
around \textit{one} of the two peaks. Hence, blindly computing $\mathbb{C}$
or $\mathcal{L}$ from this $P^{\ast }$ will not allow us to compare them
with the predictions of LGA. Indeed, it is a highly non-trivial task to
interpret $\mathcal{L}$, since a detailed understanding of the contributions
from tunneling is necessary. As for comparisons with simulations results,
the key is whether the runs are long enough to permit a good sampling of
both wells. In summary, while the LGA succeeds in capturing the essentials
of the $z$ systems here, the finite size effects are too large for good
quantitative agreement. For $z<z_{c}$ cases, we expect that the predictive
power of the LGA will improve when the zealotry is asymmetric since
tunneling events then become extremely rare~\cite{MM15,AMZ2}. \vspace{0.1cm} 
\newline
\textit{Simulation studies:} While the above methods yield exact results,
they are restricted to small systems. To study larger $N$'s, we rely on
stochastic simulations, based on running a random walker on a $S\times S$
lattice with the biased and inhomogeneous stepping probabilities~(\ref{w+},%
\ref{w-}), and performed using the Gillespie algorithm~\cite{Gillespie}. 
Using $z$'s not particularly close to either $0$ or $z_{c}$, our runs are up
to $10^{8}$ steps, for systems as large as $N=3600$. The entire trajectory
of each run is recorded, giving us the time series $\vec{n}\left( T\right) $
and so, $\vec{x}\left( t\right) $ and $\vec{\xi}\left( t\right) $. Examining
these, we find that, within 1000 time steps of starting at $\left(
0,0\right) $, the walk appears to be in a steady state. With these traces,
we can construct time averages and obtain $\left\langle \vec{x}\right\rangle 
$ and the general two-point correlation function $C_{ij}\left( \tau \right) $
in the NESS .

\begin{figure}[tbp]
\begin{center}
\includegraphics[width=3.0in, height=1.80in,clip=]{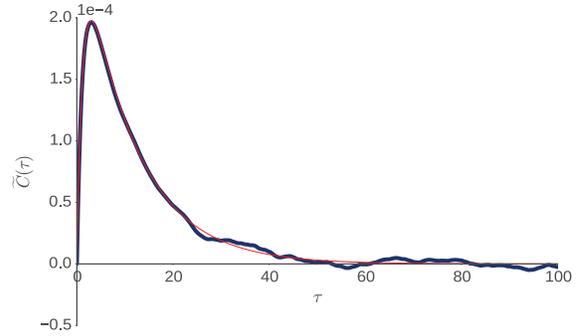}
\end{center}
\par
\vspace{-3mm}
\caption{ \textit{(Color online)}. $\tilde{C}\left( \protect\tau \right) $
vs. $\protect\tau$ in the high zealotry phase, with $N=100$, $Z=20$, and $%
S=30$: The results of simulations (black) obtained by sampling every $0.01$ time
step and averaged over 99$\times 10^6$ data points. Comparison with the LGA
expression (\protect\ref{Ltau}) with $\protect\lambda_+=0.833$ and $%
\protect\lambda_-=0.091$ (red/grey), see text.}
\label{Fig2}
\end{figure}

First, as a check, we collected data for the small systems discussed above ($%
S=30$, with $Z=20,13$). For the former, we find the truncated $C_{ij}\left(
0\right) $ to be $(C_{11},C_{12},C_{22})=\left( 1.74,1.66,3.53\right) \times
10^{-3}$. Further, we compile $\tilde{C}\left( \tau \right) $ and find the
behavior predicted in Eq.~(\ref{Ltau}), see Fig.~\ref{Fig2}. By fitting with
this form, we find $\lambda _{\pm }=0.833,0.091$ and $\mathcal{L}\cong
2.16\times 10^{-4}$. Both $C_{ij}\left( 0\right) $ and $\mathcal{L}$ are in
excellent agreement (within 0.6\%) with the exact result~\cite{AMZ_SM}. By
contrast, the LGA predictions are qualitatively acceptable (from a few \% to
the right order of magnitude). For the $Z=13$ case, the walker spends much
of its time wandering back and forth between the wells, implying that $%
P^{\ast }$ is attained. As a result, the findings for $\mathbb{C}$ are also
in excellent agreement (within 1\%) with the exact results~\cite{AMZ_SM}.
However, in this case the comparison with the LGA predictions is pointless,
since the LGA is devised for just one well.

Turning to large systems, two examples ($Z=400$ and $800$, $S=1000$) are
offered in the SM~\cite{AMZ_SM}. In the low zealotry case, the walker
remains in \textit{one well} for the entire run. Thus, it is meaningful to
compare both sets of data with MFA/LGA predictions. For $z<z_{c}$ , the
simulation results of $\left\langle \vec{x}\right\rangle $ compares well
with the MFA  $\vec{x}^{\left( \pm \right) }$. In both cases, the data for
the correlations $\left\langle n_{i}n_{j}\right\rangle _{0}$ and $%
\left\langle n_{i}n_{j}\right\rangle _{1}$ are in quantitatively good
agreement with the LGA predictions for $\mathbb{C}$ and $\mathbb{FC}$~\cite%
{AMZ_SM}. In summary, we have found that the LGA is indeed quite reliable
for large $S,Z$ (around each $\vec{x}^{\ast }$). Meanwhile, the various
methods presented in this section all point to the presence of interesting
new phenomena associated with the NESS aspect of $2q$VZ, namely, the
presence of observable quantities odd under time reversal.

\section{Summary and outlook}

In this Letter, we introduced a generalization of the $q$VMZ ($q$-voter
model with zealots)~\cite{MM15} which takes into account expected
inhomogeneities in the behavior of swing voters and the presence of zealots.
In arguably the simplest generalization of the $q$VMZ, we have just \textit{%
two} groups of swing voters, distinguished by needing a consensus of $q_{1}$
or $q_{2}$ of its neighbors to adopt their opinion. As in Ref.~\cite{MM15},
our model is characterized by two phases: When the fraction $z$ of zealots
is low, the long-time opinion distribution is bimodal whereas it is single-peaked when $z$ is above a critical value $z_c$.
However, a major and
far-reaching difference between the $q$VMZ and ours is that detailed balance
is violated here. Hence, though the qualitative features are similar to
those recently reported in \cite{MM15}, our system settles into a genuine
NESS. As a result, there are persistent probability current loops which are
odd under time reversal. We investigate these currents and some observable
manifestations thereof, in the simple but generic case $q_{1,2}=1,2$. with $%
S_{1}=S_{2}$ and $Z_{+}=Z_{-}$. Using numerical methods for small systems,
Gaussian approximation for large ones, and simulation runs for both, we
arrive at a comprehensive picture for our system. Quite remarkably, we show
that this simple model exhibits stationary microscopic current loops (see
Fig.~\ref{Fig1}), resulting in oscillations in certain macroscopic
observables (e.g., the antisymmetric part of the two-point correlation
function at unequal times, see Fig.~\ref{Fig2}). While the detailed
relationships between microscopic probability currents and macroscopic
social phenomena remain to be explored, our study clearly points to the
presence, albeit subtle, of predator-prey like oscillations. The overall
counter-clockwise flow in $\vec{K}^{\ast }$ implies that fluctuations in the 
$q_{2}$-susceptibles follow those of the other group, much like the
population of lynxs follow those of hares. We believe this stems from the
presence of \textquotedblleft leaders\textquotedblright\ and
\textquotedblleft followers\textquotedblright\ in a society. Clearly, this
study provides only the initial steps towards a systematic investigation of
multi-$q$ VM's, which are expected to display other interesting phenomena.
For this particular model, much more can be examined, e.g., finite size
effects, scaling properties near $z_{c}$, and full distributions of $%
\mathcal{L~}$\cite{SZ2014}. Beyond studying the $2q$VZ and similarly
tractable models, the goal of our long-term efforts is to explore fundamental issues
of NESS, in an attempt to formulate an overarching framework for
non-equilibrium statistical mechanics.

\acknowledgments We are grateful to B Schmittmann and J B Weiss for
enlightening discussions. This research is supported partly by a US National
Science Foundation grant: OCE-1245944. The support of the UK EPSRC and Bloom
Agency (Leeds, UK) via a CASE Studentship to AM is gratefully acknowledged
(Grant EP/L50550X/1).

\end{document}